\documentclass[prd,twocolumn,eqsecnum,longbibliography]{revtex4-1}
\usepackage{amsmath,latexsym,amssymb,mathrsfs,amsthm}
\usepackage[colorlinks=true, urlcolor=blue, citecolor=blue, linkcolor=blue, pagecolor=blue]{hyperref}
\usepackage{graphicx}

\def\be{\begin{equation}}
\def\ee{\end{equation}}
\def\bea{\begin{eqnarray}}
\def\eea{\end{eqnarray}}
\def\beq{\begin{eqnarray}}
\def\eeq{\end{eqnarray}}
\def\bas{\begin{subequations}\begin{eqnarray}}
\def\eas{\end{eqnarray}\end{subequations}}

\def\SU{\text{SU}}

\def\su{\mathfrak{su}}

\newcommand{\cC}{{\mathcal C}}

\newcommand{\bes}{\begin{eqnarray}}
\newcommand{\ees}{\end{eqnarray}}

\begin{document}
\title{\large Polymer Schwarzschild black hole: An effective metric}

\author{ J. Ben Achour$^{1}$\;\;\; F. Lamy$^{2}$\;\;\; H. Liu$^{2,3,4}$\;\;\; K. Noui$^{2,4}$\\} \vskip.2cm
\affiliation{\small\textit{ $^{1}$ Department of Physics, Beijing Normal University, Beijing 100875, China \\ $^{2}$ Laboratoire Astroparticule et Cosmologie,  Universit\'e Paris Diderot Paris 7, CNRS, 75013 Paris, France\\
$^{3}$ Centre de Physique Th\'eorique, Universit\'es d'Aix-Marseille et de Toulon, CNRS, 13288 Marseille, France \\ $^{4}$   Institut Denis Poisson, Universit\'e d'Orl\'eans, Universit\'e de Tours, CNRS,  37200 Tours, France}}

\begin{abstract}
We consider the modified Einstein equations obtained in the framework of effective spherically symmetric polymer models inspired by Loop Quantum Gravity. When one takes into account the anomaly free point-wise holonomy quantum corrections, the modification
of Einstein equations is parametrized by a function $f(x)$ of one phase space variable. We solve explicitly these equations for a
static interior black hole geometry and find the effective metric describing the trapped region, inside the black hole, for any $f(x)$. This general resolution allows to take into account a standard ambiguity inherent to the polymer regularization: namely the choice of the spin $j$ labelling the $\SU(2)$-representation of the holonomy corrections. When $j=1/2$, the function $f(x)$ is the usual sine function used in the polymer litterature. For this simple case, the effective exterior metric remains the classical Schwarzschild's one but acquires modifications inside the hole. The interior metric describes a regular trapped region and presents strong similarities with the Reissner-Nordstr\"om metric, with a new inner horizon generated by quantum effects. We discuss the gluing of our interior solution to the exterior Schwarzschild metric and the challenge to extend the solution outside the trapped region due to covariance requirement. By starting from the anomaly free polymer regularization for inhomogeneous spherically symmetric geometry, and then reducing to the homogeneous interior problem, we provide an alternative treatment to existing polymer interior black hole models which focus directly on the interior geometry, ignoring covariance issue when introducing the polymer regularization.
\end{abstract}

\maketitle

\section{Introduction}

The extraordinary recent detections of gravitational waves (GW) by the LIGO and the LIGO/Virgo collaborations
allowed us to ``hear"  black holes for the first time, a century after 
Schwarzschild predicted their existence from Einstein equations. These detections have opened a new window on 
black holes and we hope to learn much more on these fascinating astrophysical
objects in a near future. So far, the observations of GW emitted by binaries of black holes or neutron stars are in total agreement with the predictions of general relativity. However, when the GW detectors become more
 sensitive and allow probing deeper the ``very strong" gravity regime at the merger, one will possibly measure deviations from Einstein gravity. 

Perhaps, the main reason to expect gravity to be modified is the existence of singularity theorems in classical gravity. The presence of such singularity is believed to be pathological and to indicate a breakdown of the classical theory which should be modified and regularized by quantum gravity effects. However, how  quantum gravity regularizes precisely black holes singularities is still unknown simply because a complete theory of quantum gravity is still missing. Faced with such an important difficulty, one has instead proposed candidates for regular metrics with the requirements that they are non-singular modifications of the classical
black hole metric and they are physically reasonable. The Hayward \cite{Hayward:2005gi} or more recently the Planck star metrics \cite{DeLorenzo:2014pta} are typical examples. 
Hence, the regular metrics could be interpreted as effective quantum geometries.
From this point of view, it is natural to think that they could be recovered  from a semi-classical limit of a black hole quantum geometry.
In practice, this is an extremely difficult problem since it will require the development of suitable coarse-graining technics of the underlying quantum geometry, a major challenge in non-perturbative approach to quantum gravity such as Loop Quantum Gravity. %The main reason is again that
%a complete theory of quantum gravity remains elusive up to now. 

One way to circumvent this difficulty would be to construct and classify (from first principles) effective theories of quantum gravity as one does for studying in a systematic way dark energy for instance. See \cite{Bojowald:2016hgh, Carballo-Rubio:2018czn, Ghersi:2017ley} for efforts along this line.
 In that way, one could write a modified gravity action 
(or modified Einstein equations) which takes into account quantum corrections, and then study the spherically symmetric sector and look for black hole solutions. Of course, these solutions are expected to be regular and to 
predict new physical phenomena which could be in principle observable. 
In the framework of loop quantum cosmology \cite{Ashtekar:2011ni}, one knows how to construct and classify
effective quantum Friedmann equations (depending on the choice of the spin-$j$ representation which labels the holonomy corrections, as well as the choice of regularization scheme). See \cite{BenAchour:2016ajk} for details on this classification. 
It is well-known that they lead to a regular cosmology with no more initial singularity. 
However, the effective description of loop quantum black holes is much less understood, the challenge being to generalize the technics applied in LQC to the inhomogeneous black hole background. Indeed, in this inhomogeneous case, one has to make sure that the effective corrections do no generate anomalies in the algebra of first class constraints, and thus do not spoil covariance. Taking care of this potential covariance issue, one can obtain modified Einstein's equation for polymer black holes \cite{Bojowald:2015zha, Tibrewala:2013kba, Tibrewala:2012xb}. Their resolution for the simple vacuum modified Schwarzschild interior has not been investigate yet. In this letter, we fill this gap.

In the polymer framework, %the effective Einstein equations are written in a Hamiltonian form. Thus, the effective dynamics
%is totally determined by the expression of the effective Hamiltonian constraint which is regularized compared to
%the classical expression
the effective corrections are introduced at the phase space level, in the hamiltonian constraint. In the treatment of interior black holes, several regularization schemes have been developed. Models such as \cite{Ashtekar:2005qt, Bohmer:2007wi} and more recently \cite{Corichi:2015xia, Cortez:2017alh, Olmedo:2017lvt, Protter:2018tbj} make use of the homogeneity of the interior geometry to introduce a regularization very similar to cosmological polymer models. Yet, the exterior black hole geometry is inhomogeneous, and the modified Einstein's equations obtained in \cite{Ashtekar:2005qt, Bohmer:2007wi, Corichi:2015xia, Cortez:2017alh, Olmedo:2017lvt, Protter:2018tbj} hold only for the interior geometry. In this letter, we adopt a different strategy. We consider the full inhomogeneous geometry, and introduce the polymer regularization satisfying the anomaly freedom conditions of \cite{Bojowald:2015zha}, paying thus attention to the underlying covariance of the effective approach. Only after, we reduce the problem to the interior homogeneous geometry. The advantage is that we obtain one and only one set of modified Einstein's field equations valid for the whole black hole geometry (both exterior and interior regions), i.e Eq.(\ref{feq1})-(\ref{feq2}). The modified field's equation for interior region are then simply obtained by suitable gauge fixing.

Following thus the approach of \cite{Bojowald:2015zha}, the quantum corrections of the  effective Hamiltonian constraint, induced by the regularization, are parametrized by a single real valued function $f(x)$ of one phase space variable $x$. This is a consequence of the requirement that the deformed symmetry algebra (generated by the effective Hamiltonian and vectorial constraint) remains closed so that there is no anomalies. See Eq. \eqref{Hdef} and discussion below.  However, even though there is a standard choice for $f(x)$ in loop quantum gravity, 
the precise definition of the ``regularization" function $f(x)$ is in fact ambiguous.
For this reason, it is important to study 
the effective corrected Einstein equations for an arbitrary function $f(x)$, as initiated in \cite{Bojowald:2015zha}. 
 
In this letter, we consider the effective theory introduced in \cite{Bojowald:2015zha} and we solve explicitly the effective Einstein equations for static spherically symmetric interior space-times. More precisely, we
focus on the static region inside the horizon, where quantum gravity effects are supposed to become important, and we find an explicit form
of the effective metric in this region for an arbitrary deformation function $f(x)$.  Surprisingly, the effective metric can be simply 
expressed in terms of  $f(x)$, and then we can easily deduce the conditions for the black hole to be
non-singular as one wishes. We apply our result to the case where $f(x)$ is the standard deformation function used in loop quantum gravity
\eqref{loopf}, and we show that the black hole presents strong similarities with the Reissner-Nordstr\"om space-time. The interior effective geometry inherits an inner horizon due to the non perturbative quantum gravity effects. Equipped with this new interior effective geometry, we explore then the possibility to extend our black hole solution to the whole space-time (outside the trapped region) and we discuss the challenge to perform coordinate transformation in this model with deformed covariance. Finally, we apply the strategy developed in \cite{Bojowald:2018xxu} to obtain a well defined invariant line element under the deformed symmetry and we show that the main novelty is a transition between Lorentzian to Euclidean signature deep inside the interior region. 

\section{Covariant polymer phase space regularization}
Let us first present the effective Einstein equations obtained in loop quantum gravity
for spherically symmetric black holes and justify our choice of regularization. 

%As we are interested in a Hamiltonian formulation of the theory
%in the region inside the black hole (where quantum
%gravity effects become important), 
We start with an ADM parametrization of the metric
\be
\label{ADM}
ds^2 = - N^2\,  dt^2 + q_{rr}\, ( dr + N^r dt )^2 + q_{\theta\theta} \, d\Omega^2 
\ee
where each function $N$, $N^r$, $q_{rr}$ and $q_{\theta\theta}$ depends on the radial and time coordinates $(t,r)$,
and $d\Omega^2$ is the metric on the unit two-sphere. 
Following the Ashtekar-Barbero construction, it is more convenient to express the metric components $q_{rr}$ and $q_{\theta \theta}$ 
in terms of the components of the electric field $E^r$ and $E^\phi$ as follows
\be
q_{rr} \equiv \frac{(E^{\phi})^2}{E^r} \, ,\qquad q_{\theta\theta} \equiv E^r  \, .
\ee
Hence, the phase space is parametrized by two pairs of conjugate fields defined by the Poisson brackets
\be
\label{phase space}
\{ K_{\phi} (r), E^{\phi}(s)\} =  \delta(r -s) \, , \, \{ K_r (r), E^r (s) \} = 2 \delta(r -s) \;\;\;\;\;\;\;\;
\ee
where we have fixed for simplicity the Newton constant and the Barbero-Immirzi parameter to $1$.
The variables $K_{\phi}$ and $K_r$ are $\su(2)$ connections.

As usual, the lapse function $N$ and the shift vector $N^r$ are Lagrange
multipliers which enforce respectively the Hamiltonian and vectorial constraints,
\bea
H & = &  \frac{E^{\phi}}{2\sqrt{E^r}} (1+ K^2_{\phi}  - \Gamma^2_{\phi}) +  \sqrt{E^r} (K_{\phi} K_r  +  \partial_r\Gamma_{\phi}) \;\;\; \;\;\; \label{Hconstraint}\\
V & = & 2 E^{\phi} \partial_r K_{\phi} - K_r \partial_r E^r \, , \label{Vconstraint}
\eea
where $\Gamma_{\phi} \equiv - {\partial_r E^r}/{2E^{\phi}}$ is linked to the Levi-Civita connection.
These constraints are first class, they generate diffeomorphisms restricted to spherically symmetric space-times, and they 
satisfy the closed Poisson algebra
\begin{align}
& \{ H[N], V[N_1^r]\} = - H[N_1^r  \partial_r N]  \, ,\label{HV}\\
& \{ V[N^r_1], V[N^r_2]\} = V [N_1^r \partial_r N^r_2 - N_2^r \partial_r N^r_1] \, ,\label{VV}\\
& \{ H [N_1], H[N_2]\} = V [q^{rr} (N_1 \partial_r N_2 - N_2 \partial_rN_1)] \, ,\label{HH}
\end{align}
where $H[N]$ and $V[N_1^r]$ are the smeared constraints. 

%Quantizing this system means to promote the phase space variables \eqref{phase space} into non-commutative operators, 
%to find unitary irreducible representations of this operators algebra, and finally to compute solutions of the quantum constraints.  
%In practice, this program is difficult to apply and one encounters ambiguities in constructing the quantum constraints. Several quantizations of the (holonomy corrected) phase space was proposed in \cite{Bojowald:2004af, Bojowald:2004ag, Bojowald:2005cb, Gambini:2013ooa}. 

In this letter, we focus on the effective dynamic obtained from the anomaly free loop regularization which is introduced prior quantization. Concretely, we keep the phase space parametrization
 \eqref{phase space} unchanged and we modify the expression of the constraint \eqref{Hconstraint}. As we
consider solely point-wise holonomy corrections of $K_{\phi}$ here, only the dependency of the Hamiltonian constraint on $K_\phi$ is modified according to
\be
\label{Hdef}
H= \frac{E^{\phi}}{2\sqrt{E^r}} [1+ {f}(K_{\phi})  - \Gamma^2_{\phi}] +  \sqrt{E^r} [ g(K_{\phi}) K_r  +  \partial_r\Gamma_{\phi}] \;\;\;\;\;\;\;\;
\ee
where the functions $f$ and $g$ are not fixed yet. The requirement of anomaly freedom of the Dirac's algebra requires then that 
\be
g(x) = f'(x)/2
\ee 
In that case, the Poisson bracket
between Hamiltonian constraints \eqref{HH} is deformed according to
\be
\label{defHH}
 \{ H [N_1], H[N_2]\} = V [\beta(K_\phi) \, q^{rr} (N_1 N'_2 - N_2 N'_1)] \;\;\;\;\;\;
\ee
where the deformation function $\beta(K_{\phi})$ is given by
\be
\beta(x)=f''(x)/2
\ee
as initially derived in \cite{Bojowald:2015zha, Tibrewala:2013kba}. Such a deformation is a generic feature of holonomy corrected symmetry reduced models of gravity \cite{Bojowald:2016itl}. The other two brackets \eqref{HV} and \eqref{VV} are unchanged. Moreover, $K_r$ is not modified in our regularization since it can be completely remove from the scalar constraint by a simple redefinition of the constraints, as shown in \cite{Gambini:2013ooa}. Consequently, the regularization of $K_r$ doesn't play any role in the classical regularization and can be safely ignored at this step. The holonomies of $K_r$ will nevertheless be crucial in the quantum theory when introducing the one dimensional spin network defining the kinematical Hilbert space. See \cite{Gambini:2013ooa} for more details. 

Finally, our regularization is restricted to the $\mu_0$-scheme, as in \cite{Gambini:2013ooa}, since introducing holonomy corrections within the $\bar{\mu}$-scheme, i.e $K_{\phi} \rightarrow f(K_{\phi}, E^x)$, and requiring at the same time the anomaly freedom of the effective Dirac's algebra generates inconsistencies as shown in \cite{Tibrewala:2012xb}. Therefore, the standard improved dynamics used in polymer cosmological models cannot be generalized as it stands to such inhomogeneous spherically symmetric polymer models. See \cite{Corichi:2015xia, Cortez:2017alh, Olmedo:2017lvt, Protter:2018tbj} for a recent alternative strategy. This concludes our justifications for our classical regularization of the phase space.

%{\Jib Therefore, we can obtained an anomaly free, (albeit deformed) regularization of our phase space. Additionally, the existence of an equivalent formulation of the phase space where the new scalar constraint $\cC$ does not depend on $K_r$  suggests that effective corrections of this component should not play a crucial role in the resulting modified black hole solution we are looking for. We will therefore ignore such quantum corrections, contrary to alternative models of black hole interior which include such corrections \cite{Ashtekar:2005qt, Campiglia:2007pb, Corichi:2015xia, Olmedo:2017lvt}. Notice moreover that this choice of regularization allows to work with a set of modified constraints which is consistent both in the homogeneous inside region as well as in the inhomogeneous outside region, contrary to \cite{}. This choice coincides with the one of Gambini and Pullin in \cite{Gambini:2013ooa}, where the whole space-time was considered. Finally, we stress that our regularization does not include the so called $\bar{\mu}$-scheme, but this restriction is not special to our treatement, since it is shared by any other black hole models of the litterature so far \cite{Ashtekar:2005qt, Campiglia:2007pb, Corichi:2015xia, Olmedo:2017lvt}. Moreover, the implementation of the $\bar{\mu}$-scheme in inhomogeneous spherically symmetric models is out of reach with the standard technics, as discussed in \cite{Tibrewala:2013kba}. Having justify our choice of regularization, we can now investigate the effective field equations.}

\section{Effective Einstein's equations}

Hence, as it was emphasized in the introduction, the regularization induced by holonomy corrections inspired from loop quantum gravity is parametrized by the sole function $f(x)$. The explicit expression of this effective correction remains ambiguous. 
Nonetheless, as we require naturally that $f(x)$ reproduces the classical behavior in the low 
curvature regime, we must have $f(x) \approx x^2$ when $x \ll 1$. In the literature, the usual choice is
\bea
\label{loopf}
f(x) = \frac{\sin^2(\rho x)}{\rho^2} \, ,
\eea
where $\rho$ is a deformation real parameter that tends to zero at the classical limit. The presence of a trigonometric function is reminiscent
from the $SU(2)$ gauge invariance in loop quantum gravity: roughly, one replaces the ``connection" variable $K_\phi$ by a point-wise ``holonomy-like" 
variable $\sin(\rho K_\phi)/\rho$. Note that (\ref{loopf}) is associated to the computation of the regularization of the connection (or its curvature) in term of holonomies within the $j=1/2$ fundamental representation of $\SU(2)$. Yet, one could obtain more complicated trigonometric functions by evaluating this regularization in another $j$-representation of $\SU(2)$, as done for polymer cosmological models in \cite{BenAchour:2016ajk}. Therefore, keeping f(x) general in our resolution allows to keep track of this ambiguity of the polymer regularization. 
 
Now, we have all the ingredients to compute the effective Einstein equations for deformed spherically symmetric space-times.
They are given by the Hamilton equations 
\bea \dot F= \{F,H[N] + V[N^r]\} \, ,\eea
for $F$ being one of the four phase space variables 
\eqref{phase space}. The time evolutions of the electric field components simply read
\begin{align}
\label{feq1}
& \dot{E}^r = N \sqrt{E^r} f'\!(K_{\phi}) + N^r \partial_r E^r  \, ,\\
\label{feq2}
& \dot{E^{\phi}} = \frac{N}{2} \left[\sqrt{E^r} K_r f''\!(K_{\phi}) + \frac{E^{\phi}}{\sqrt{E^r}} f'\!(K_{\phi})\right] + \partial_r(N^r E^{\phi})  \, .
\end{align}
The expression of $\dot K_\phi$ is more involved and thus we do not report it here.
The component $K_r$ can be obtained by solving the Hamiltonian constraint \eqref{Hdef}. 

\subsection{Outside the black hole}

Note that for $N^r=0$ and static geometry, and upon using the standard loop effective corrections (\ref{loopf}), equation (\ref{feq1}) implies that the angular extrinsic curvature is quantized as
\be
K_{\phi} = \frac{n \pi }{2\rho} \qquad \text{with} \qquad  n \in \mathbb{N}
\ee 
Hence, for $n\neq 0$, the resulting geometry has a divergent extrinsic curvature $K_{\phi}$ in the semi-classical limit, i.e when $\rho \rightarrow 0$.
It implies that outside the hole, the only consistent inhomogeneous static solution is the classical Schwarzschild's one, i.e 
\be
K_{\phi}=0
\ee
corresponding to $n=0$.
Therefore, the effective loop corrections introduced above do not allow to have a modified Schwarzschild geometry outside the hole when looking for a static exterior solution. This shortcoming is intimately related to the lack of a proper $\bar{\mu}$-scheme in the present regularization. It is expected that once a fully consistent $\bar{\mu}$-scheme will be implemented, i.e with a polymer scale $\rho(E^x)$ running with the geometry, potential modifications of the exterior geometry could show up.

\subsection{Inside the black hole: static ansatz}
We turn now to the interior problem. We are interested in solving these equations inside a ``static" black hole. As the role of the variables $r$ and $t$ changes when one crosses
the horizon, this corresponds to considering time-dependent fields only. In that case, the effective Einstein equations dramatically simplify and read
\begin{align}
& \dot{E}^r = N \sqrt{E^r} f'\!(K_{\phi})  \, , \label{eqEr}\\
& \dot{E^{\phi}} = \frac{N}{2} \left[ \sqrt{E^r} K_r f''\!(K_{\phi}) + \frac{E^{\phi}}{\sqrt{E^r}} f'\!(K_{\phi}) \right] \, ,\label{eqEphi}\\
& \dot{K}_{\phi} =  - \frac{N}{2 \sqrt{E^r} }\left[ 1 + f(K_{\phi})\right] \, ,\label{eqKphi}
\end{align} 
together with the Hamiltonian constraint
\bea
\label{Hamt}
 f'\!(K_\phi) E^r K_r+ [1+ f(K_\phi)] E^\phi \; = \; 0 \, ,
\eea
from where we easily get the dynamics of  ${K}_{r}$.

\subsubsection{General algorithm}

Now, we are going to solve these equations explicitly for any function $f$. As we are going to show, it is very convenient to fix the lapse 
function $N(t)$ (by a gauge fixing) such that
\bea
\label{gauge}
N \; f'\!(K_\phi) \; = \; 2 \, .
\eea
In that case, the equation \eqref{eqEr} for $E^r$ decouples completely from the other variables and can be easily integrated to
\bea
\label{Er}
E^r(t) \; = \; t^2 + a \, ,
\eea
where $a$ is an integration constant that we fix to $a=0$ (in order to recover the Schwarzschild solution at the classical limit). 
Another important consequence of the gauge choice
\eqref{gauge} is that the equation \eqref{eqKphi} for $K_\phi$ also decouples and takes the very simple form
\bea
\frac{f'\!(K_\phi) }{1+f(K_\phi)}  \, \dot K_\phi \; = \; - \frac{1}{t} \, . 
\eea
It can be immediately integrated to the form
\bea
f(K_\phi) = \frac{r_s}{t} - 1 \, ,
\eea
where $r_s$ is an integration constant with the dimension of a length. As we are going to see later on, 
$t=r_s$ corresponds to the location of the black hole (outer) horizon. Hence, $K_\phi$ is easily obtained  by inverting the
function $f(x)$. Indeed, when $f$
is monotonous, it admits a global reciprocal function $f^{-1}$, otherwise the reciprocal function is defined locally. Then, the expression of $E^\phi$ follows immediately. Indeed, if one substitutes $K_r$ from \eqref{Hamt} into 
\eqref{eqEphi}, one obtains the following equation for $E^\phi$
\bea
\frac{\dot E^\phi}{E^\phi} = \frac{1}{t} \left( 1 - \frac{[1+f(K_\phi)] f''\!(K_\phi)}{[f'\!(K_\phi)]^2}\right)\, ,
\eea
which can be easily integrated to
\bea
\label{Ephi}
E^\phi \; = \; b \frac{f'\!(K_\phi)}{1+ f(K_\phi)} \, ,
\eea
where $b$ is a new integration constant that will be fixed later. The remaining variable $K_r$ is given immediately
from the Hamiltonian constraint \eqref{Hamt} together with \eqref{Er} and \eqref{Ephi}. Hence, we have integrated explicitly
and completely the modified Einstein equations in the region inside a ``static" spherically symmetric black hole where the effective
metric  is 
\bea
\label{solmet}
ds^2= -\frac{1}{F(t)} dt^2 + \left(\frac{2b}{r_s}\right)^2 F(t) dr^2 + t^2 d\Omega^2 \, ,
\eea
with $F(t)$ related to $f(x)$ by
\bea
\label{relFf}
F(t) = \frac{1}{4}\left[f' \circ f^{-1}(\frac{r_s}{t}-1)\right]^2 =
\left[2\frac{d f^{-1}}{dx} (\frac{r_s}{t}-1) \right]^{-2} \;\;\;\;\;\; .
\eea
In the region where $t \approx r_s$, quantum gravity effects are negligible and the metric should reproduce the Schwarzschild metric. 
We see immediately in \eqref{solmet} that a necessary condition for this to be the case is that 
\be
2b=r_s
\ee
This fixes the constant $b$. Furthermore, in such a regime, we know that $f(x) \approx x^2$, then $f^{-1}(x)\approx \sqrt{x}$, hence 
\be
F(t) \approx \vert r_s/t - 1 \vert
\ee
As a consequence,
we recover the expected classical metric with $r_s$ being the Schwarzschild radius. However, while the metric smoothly matches the Schwarzschild metric at the outer horizon, the extrinsic  curvature does not, leaving a gluing which is not $\cC^1$.

\subsubsection{Inverse problem}

Before studying concrete examples, let us consider a converse situation where a deformed metric $g_{\mu\nu}$ of the form 
\eqref{solmet} is given. Then, one asks the question whether one can find a deformation function $f(x)$ such that the deformed metric
$g_{\mu\nu}$ is a solution of the effective Einstein equations. The answer is positive and $f(x)$ can be obtained immediately by inverting
the relation \eqref{relFf} between $F(t)$ and $f(x)$ as follows
\bea
f^{-1}(x) \; = \; \frac{1}{2} \int_0^x {du}\,{{\left\vert F\left( \frac{r_s}{1+u}\right)\right\vert}}^{-1/2} \, .
\eea
As the function $f^{-1}(x)$ is monotonic, one can invert this relation and define the deformation function $f(x)$ 
without ambiguity. This can be done for the Hayward metric for instance, even though in that case $f^{-1}(x)$ is defined
as an integral, and thus $f(x)$ is implicit.

\subsection{Example: the standard $j=1/2$ sine correction} To illustrate this result, let us consider some interesting physical situations. 
First, the case where there is no quantum deformation corresponds to $f(x)=x^2$. As we have just said above, we 
recover immediately the Schwarzschild metric. 

Then, let us study the more interesting case where $f(x)$ is the usual function considered in polymer black hole models \eqref{loopf}. 
In that case, the reciprocal function is 
\bea
f^{-1}(x)= \frac{\arcsin(\rho \sqrt{x})}{\rho} \, ,
\eea
which is defined for $x \leq 1/\rho^2$ only. As a consequence, the effective metric for a black hole
is of the form \eqref{solmet} with
\bea
F(t) \; = \; \left( \frac{r_s}{t} -1\right) \left( 1+ \rho^2 - \rho^2 \frac{r_s}{t}\right) \,,
\eea
which is defined for $t \leq r_s$ a priori. At this point, we can make several interesting remarks. First, one recovers the Schwarzschild metric,
when $t$ approaches $r_s$. Then, in addition to the usual outer horizon (located at $t=r_s$), the metric has an inner horizon located at
\be 
t= \frac{\rho^2r_s}{1+\rho^2}
\ee
The computation of the Ricci and Kretschmann scalars shows that there is no curvature singularity inside the trapped region. One can naturally extend this solution outside the trapped region by using  a generalized advanced time coordinate $v$ such that
\be
dv=dr+dt/F(t)
\ee
and
\be
ds^2 = F(t) dv^2-2dvdt + t^2 d\Omega^2
\ee
The metric and inverse metric are regular when $F(t)=0$. This allows to define the expansion of null radial outgoing geodesics, leading to 
\be
\theta_+=F(t)/t
\ee 
Hence the zeros of $F$ correspond to the locus of the horizons (inner and outer), and the region comprised between them is trapped. The Ricci scalar 
\be
R \approx -2\rho^2/t^2
\ee 
diverges at $t=0$, which is the locus of a timelike singularity as in Reissner-Nordstr\"om's (RN) black hole. Our metric is actually very similar to this solution, and leads to the same Penrose diagram. However, a main difference is that an outer horizon (at $t=r_s$) is always present in our geometry, while naked singularities appear for super-extreme RN black holes. 

In the end, this naive extension is not satisfactory since it does not allow recovering Schwarzschild's solution in the classical region ($r \gg r_s$), except if the parameter $\rho$ becomes $r$-dependent and tends to zero, 
which would drastically modify the equations of motion \cite{Tibrewala:2012xb}.

\subsection{Signature change from covariance}

Moreover, while the extension of the metric outside the trapped region is natural for a standard RN solution, it is not clear whether the extension is  allowed or not in our context. The reason is that the deformation of the Hamiltonian constraint (\ref{defHH}) modifies the invariance of the effective theory under time reparametrizations. Then,  if we believe that such a deformed symmetry is the right one and it is no longer given
by usual diffeomorphisms (what can be discussed), 
we could not perform an arbitrary time redefinition  as we did to extend the metric outside the trapped region. The 
deformed symmetry has recently been analyzed in great details in  \cite{Bojowald:2018xxu}. It was realized 
that the effective metric which is invariant under these deformed transformations is slightly different from \eqref{ADM}
where $N$ has to be rescaled according to
\bea
\label{beta}
N^2 \longrightarrow \, \, \beta(K_\phi) \, N^2 \, ,
\eea
where $\beta$, which has been introduced in \eqref{defHH}, is explicitly given, in our case, as a function of time
by 
\be
\beta(t) \equiv \beta(K_\phi(t)) = 1 - 2 \rho^2 \left( \frac{r_s}{t} - 1\right) \, .
\ee 
With this new invariant metric, the function $g_{tt}(t)$ acquires a zero in the trapped region which corresponds to a transition between a lorentzian and an euclidean signature within the trapped region at
\be 
t = \frac{2\rho^2 r_s}{1+2\rho^2}
\ee
Such a transition was studied in more detail in \cite{Bojowald:2018xxu}. Starting from another gauge choice, namely $K_{\phi}= \pi/(2\rho)$, and solving the field equations (\ref{eqEr}-\ref{Hamt}) deep inside the black hole, it was shown that the geometry is regular. Yet, the lorentzian to euclidean transition rises new difficulties concerning for instance the fate of matter inside this trapped region, since the standard evolution equations become elliptic \cite{Bojowald:2014zla}. Similar aspects were encountered in the context of the perturbations analysis in loop quantum cosmology known as the deformed algebra approach \cite{Barrau:2014maa}.

\section{Discussion} 

In this letter, we have solved explicitly a large class of modified Einstein equations arising in the effective polymer approach to black holes. We have adopted a different strategy than existing interior Schwarzschild models such as \cite{Corichi:2015xia, Cortez:2017alh, Olmedo:2017lvt, Protter:2018tbj}. We first consider the full polymer regularization of the inhomogeneous geometry consistent with covariance, and only then reduce the problem to the interior homogeneous geometry. By doing this, we ensures that the regularization of the hamiltonian constraint does not generate any anomalies and thus, that we still have the right number of degrees of freedom at the effective level. This point is ignored in \cite{Corichi:2015xia, Cortez:2017alh, Olmedo:2017lvt, Protter:2018tbj} and the regularization introduced in these models is different, since there are no anomaly freedom condition to constrain it. Consequently, the effective metric obtained in this letter and the one presented in \cite{Corichi:2015xia, Cortez:2017alh, Olmedo:2017lvt, Protter:2018tbj} are very different. 

Focusing on the usual deformation considered in polymer models studied by Gambini and Pullin, we have found a black hole (interior) solution whose structure shows strong similarities with the Reissner-Nordstr\"om black hole. The main novelty due to the quantum gravity effect is the appearance of an inner horizon, while the expected Schwarzschild solution is recovered when one approaches the outer horizon, albeit not smoothly. This last point is a consequence of the lack of a proper $\bar{\mu}$-scheme regularization in the Gambini-Pullin model.

Strictly speaking, we obtained a solution only inside in a trapped region, valid for $t \in [t_{-}, t_{+}]$ and the question of its extension in the whole space-time deserves to be study carefully. In particular, we see that the naive extension outside the trapped surface  does not allow recovering Schwarzschild's solution in the classical region ($r \gg r_s$), except if the parameter $\rho$ becomes $r$-dependent and tends to zero. This underlines the limitation of the current model to have a consistent semi-classical limit. A generalization of the current regularization is required to account for a $\overline{\mu}$-scheme, as already emphasized in \cite{Tibrewala:2012xb} and more recently in \cite{BenAchour:2017ivq}. See also \cite{BenAchour:2017jof} for a more recent proposal including such $\overline{\mu}$-scheme in polymer black holes using self dual variables.

Our results open interesting theoretical and phenomenological directions to follow. 
First, it would be  interesting to include additional effective corrections such as the triad corrections affecting the intrinsic geometry, which are
usually  considered separately. An even more challenging step is to go beyond static geometries and study dynamical black holes, an open issue up to now in the polymer framework. See \cite{BenAchour:2017jof, Campiglia:2016fzp} for recent proposals in this direction. This is particularly important for understanding Hawking radiation as well as quantum gravitational collapse and eventually bouncing and black hole to white hole transitions scenarios. Finally, it would be interesting to use this model to investigate possible quantum gravity modifications of the structure inside astrophysical objects. We plan to address these important questions in the futur. \\

\textit{Acknowledgements.} J.BA would like to thank S. Brahma for numerous discussions and for having shared with us
their results \cite{Bojowald:2018xxu} prior to publication. We would like to thank E. Livine for his interesting remarks. This work was supported by the National Science Foundation of China, Grant No.11475023 and No.11875006 (J. BA).

\end{document}